%% file: main.tex
\documentclass[12pt,a4paper]{article}
\pdfoutput=1
\usepackage{ifthen} 
\usepackage{subfigure}
\newboolean{pdflatex}
\setboolean{pdflatex}{true} 

\newboolean{articletitles}
\setboolean{articletitles}{true} 

\newboolean{uprightparticles}
\setboolean{uprightparticles}{false} 

\input{preamble}
\usepackage{longtable} 
\usepackage{caption}
\usepackage{breqn}
\usepackage[normalem]{ulem}

\usepackage{xcolor}
\begin{document}
\renewcommand{\thefootnote}{\fnsymbol{footnote}}
\setcounter{footnote}{1}

\input{title}


\renewcommand{\thefootnote}{\arabic{footnote}}
\setcounter{footnote}{0}


\pagestyle{plain} 
\setcounter{page}{1}
\pagenumbering{arabic}


\input{introduction}

\input{method}

\input{model_comp}
\input{observables}

\input{lepton_universality}

\input{conclusions}

\section*{Acknowledgements}
We would like to thank C.~Bobeth, D.~van Dyk, and J.~Virto for their help in obtaining their predictions of the charm correlator and for explaining in detail their model. We would also like to thank J.~Matias, A.~Khodjamirian and R.~Zwicky for helpful discussions. Many thanks to  S.~Harnew, C.~Langenbruch, S.~Maddrell-Mander, J.~Rademacker, M.-H.~Schune  and N.~Skidmore for their corrections to the text. GP acknowledges support from the UK Science and Technology Facilities Council (STFC) from the grant ST/N503952/1, TB acknowledges support from the Royal Society (United Kingdom) and PO acknowledges support from the Swiss National Science Foundation under grant number BSSGI0\_155990.
\clearpage
\input{appendix}
\clearpage
\bibliographystyle{LHCb}
\bibliography{main}

\end{document}

%% file: preamble.tex
\usepackage{microtype}
\usepackage{lineno}  
\usepackage{xspace} 

\usepackage{graphicx}  
\usepackage{color}
\usepackage{colortbl}
\graphicspath{{./figs/}} 

\usepackage{amsmath} 
\usepackage{amssymb}
\usepackage{amsfonts}
\usepackage{upgreek} 

\newcommand*\patchAmsMathEnvironmentForLineno[1]{%
\expandafter\let\csname old#1\expandafter\endcsname\csname #1\endcsname
\expandafter\let\csname oldend#1\expandafter\endcsname\csname
end#1\endcsname
 \renewenvironment{#1}%
   {\linenomath\csname old#1\endcsname}%
   {\csname oldend#1\endcsname\endlinenomath}%
}
\newcommand*\patchBothAmsMathEnvironmentsForLineno[1]{%
  \patchAmsMathEnvironmentForLineno{#1}%
  \patchAmsMathEnvironmentForLineno{#1*}%
}
\AtBeginDocument{%
\patchBothAmsMathEnvironmentsForLineno{equation}%
\patchBothAmsMathEnvironmentsForLineno{align}%
\patchBothAmsMathEnvironmentsForLineno{flalign}%
\patchBothAmsMathEnvironmentsForLineno{alignat}%
\patchBothAmsMathEnvironmentsForLineno{gather}%
\patchBothAmsMathEnvironmentsForLineno{multline}%
}

\usepackage{hyperref}    
\usepackage[all]{hypcap} 

\input{lhcb-symbols-def} 
\input{kmm-symbols-def}

\usepackage{cite} 
\usepackage{mciteplus}

%% file: lhcb-symbols-def.tex
\def\lhcb {\mbox{LHCb}\xspace}
\def\ux85 {\mbox{UX85}\xspace}

\ifthenelse{\boolean{uprightparticles}}%
{

 \def\Pmu         {\ensuremath{\upmu}\xspace}

 \def\Ppi         {\ensuremath{\uppi}\xspace}

 \def\Ppsi        {\ensuremath{\uppsi}\xspace}

 \def\PDelta      {\ensuremath{\Delta}\xspace}                 
 \def\PXi      {\ensuremath{\Xi}\xspace}                 
 \def\PLambda      {\ensuremath{\Lambda}\xspace}                 
 \def\PSigma      {\ensuremath{\Sigma}\xspace}                 
 \def\POmega      {\ensuremath{\Omega}\xspace}                 
 \def\PUpsilon      {\ensuremath{\Upsilon}\xspace}                 
 
 \def\PB      {\ensuremath{\mathrm{B}}\xspace}                 
                  
 \def\PD      {\ensuremath{\mathrm{D}}\xspace}

 \def\PJ      {\ensuremath{\mathrm{J}}\xspace}                 
 \def\PK      {\ensuremath{\mathrm{K}}\xspace}

 \def\Pb      {\ensuremath{\mathrm{b}}\xspace}

 \def\Pi      {\ensuremath{\mathrm{i}}\xspace}

 \def\Ps      {\ensuremath{\mathrm{s}}\xspace}

}
{

 \def\Pmu         {\ensuremath{\mu}\xspace}

 \def\Ppi         {\ensuremath{\pi}\xspace}

 \def\Ppsi        {\ensuremath{\psi}\xspace}                 
                  
 \mathchardef\PDelta="7101
 \mathchardef\PXi="7104
 \mathchardef\PLambda="7103
 \mathchardef\PSigma="7106
 \mathchardef\POmega="710A
 \mathchardef\PUpsilon="7107
                  
 \def\PB      {\ensuremath{B}\xspace}                 
                  
 \def\PD      {\ensuremath{D}\xspace}

 \def\PJ      {\ensuremath{J}\xspace}                 
 \def\PK      {\ensuremath{K}\xspace}

 \def\Pb      {\ensuremath{b}\xspace}

 \def\Pi      {\ensuremath{i}\xspace}

 \def\Ps      {\ensuremath{s}\xspace}

}

\def\mup        {\ensuremath{\Pmu^+}\xspace}
\def\mun        {\ensuremath{\Pmu^-}\xspace} 
\def\mumu       {\ensuremath{\Pmu^+\Pmu^-}\xspace}

\def\ellm       {\ensuremath{\ell^-}\xspace}
\def\ellp       {\ensuremath{\ell^+}\xspace}

\def\squark    {\ensuremath{\Ps}\xspace}

\def\bquark    {\ensuremath{\Pb}\xspace}

\def\pion  {\ensuremath{\Ppi}\xspace}
\def\piz   {\ensuremath{\pion^0}\xspace}

\def\pim   {\ensuremath{\pion^-}\xspace}

\def\kaon  {\ensuremath{\PK}\xspace}
  \def\Kbar  {\kern 0.2em\overline{\kern -0.2em \PK}{}\xspace}
\def\Kb    {\ensuremath{\Kbar}\xspace}
\def\Kz    {\ensuremath{\kaon^0}\xspace}
\def\Kzb   {\ensuremath{\Kbar^0}\xspace}
\def\KzKzb {\ensuremath{\Kz \kern -0.16em \Kzb}\xspace}
\def\Kp    {\ensuremath{\kaon^+}\xspace}
\def\Km    {\ensuremath{\kaon^-}\xspace}

\def\KpKm  {\ensuremath{\Kp \kern -0.16em \Km}\xspace}
\def\KS    {\ensuremath{\kaon^0_{\rm\scriptscriptstyle S}}\xspace} 
 
\def\Kstarz  {\ensuremath{\kaon^{*0}}\xspace}
\def\Kstarzb {\ensuremath{\Kbar^{*0}}\xspace}

  \def\Dbar    {\kern 0.2em\overline{\kern -0.2em \PD}{}\xspace}
\def\D       {\ensuremath{\PD}\xspace}

\def\Dz      {\ensuremath{\D^0}\xspace}
\def\Dzb     {\ensuremath{\Dbar^0}\xspace}
\def\DzDzb   {\ensuremath{\Dz {\kern -0.16em \Dzb}}\xspace}
\def\Dp      {\ensuremath{\D^+}\xspace}
\def\Dm      {\ensuremath{\D^-}\xspace}

\def\DpDm    {\ensuremath{\Dp {\kern -0.16em \Dm}}\xspace}

\def\B       {\ensuremath{\PB}\xspace}
\def\Bbar    {\ensuremath{\kern 0.18em\overline{\kern -0.18em \PB}{}}\xspace}
\def\Bb      {\ensuremath{\Bbar}\xspace}

\def\Bub     {\ensuremath{\B^-}\xspace}

\def\Bd      {\ensuremath{\B^0}\xspace}
\def\Bs      {\ensuremath{\B^0_\squark}\xspace}

\def\Bdb     {\ensuremath{\Bbar^0}\xspace}

\def\jpsi     {\ensuremath{{\PJ\mskip -3mu/\mskip -2mu\Ppsi\mskip 2mu}}\xspace}
\def\psitwos  {\ensuremath{\Ppsi{(2S)}}\xspace}

  \def\Y#1S{\ensuremath{\PUpsilon{(#1S)}}\xspace}

\def\Lbar {\ensuremath{\kern 0.1em\overline{\kern -0.1em\PLambda}}\xspace}

\newcommand{\decay}[2]{\ensuremath{#1\!\to #2}\xspace}         

\def\to                 {\ensuremath{\rightarrow}\xspace}

\def\qsq       {\ensuremath{q^2}\xspace}

\def\CP                {\ensuremath{C\!P}\xspace}

\def\BdToKstmm    {\decay{\Bd}{\Kstarz\mup\mun}}

\def\BdbToKstmm   {\decay{\Bdb}{\Kstarzb\mup\mun}}

\def\BdbToKstll   {\decay{\Bdb}{\Kstarzb\ellp\ellm}}

\def\BdbToPhiKst {\decay{\Bdb}{\phi\Kstarzb}}

\def\BdbToRhoKst {\decay{\Bdb}{\rho^0\Kstarzb}}
\def\BdToPsi2sKst  {\decay{\Bd}{\psitwos\Kstarz}}
\def\BdbToPsi2sKst  {\decay{\Bdb}{\psitwos\Kstarzb}}
\def\BdbToPsiTwosKst  {\decay{\Bdb}{\psitwos\Kstarzb}}
\def\BdbToPsiTwosKPi  {\decay{\Bdb}{K^-\pi^+\psitwos}}
\def\BdbToJPsiKst {\decay{\Bdb}{\jpsi\Kstarzb}}
\def\BdbToJPsiKPi {\decay{\Bdb}{K^-\pi^+\jpsi}}

\def\BuToKmm    {\decay{\Bub}{\Km\mup\mun}}
\def\BuToPimm    {\decay{\Bub}{\pim\mup\mun}}

\def\AT#1     {\ensuremath{A_{\mathrm{T}}^{#1}}\xspace}           

\def\ctl       {\ensuremath{\cos{\theta_l}}\xspace}
\def\ctk       {\ensuremath{\cos{\theta_K}}\xspace}

\def\C#1      {\ensuremath{\mathcal{C}_{#1}}\xspace}                       
\def\Cp#1     {\ensuremath{\mathcal{C}_{#1}^{'}}\xspace}                    
\def\Ceff#1   {\ensuremath{\mathcal{C}_{#1}^{\mathrm{(eff)}}}\xspace}        
\def\Cpeff#1  {\ensuremath{\mathcal{C}_{#1}^{'\mathrm{(eff)}}}\xspace}       
\def\Ope#1    {\ensuremath{\mathcal{O}_{#1}}\xspace}                       
\def\Opep#1   {\ensuremath{\mathcal{O}_{#1}^{'}}\xspace}                    

\newcommand{\tev}{\ensuremath{\mathrm{\,Te\kern -0.1em V}}\xspace}
\newcommand{\gev}{\ensuremath{\mathrm{\,Ge\kern -0.1em V}}\xspace}
\newcommand{\mev}{\ensuremath{\mathrm{\,Me\kern -0.1em V}}\xspace}
\newcommand{\kev}{\ensuremath{\mathrm{\,ke\kern -0.1em V}}\xspace}
\newcommand{\ev}{\ensuremath{\mathrm{\,e\kern -0.1em V}}\xspace}
\newcommand{\gevc}{\ensuremath{{\mathrm{\,Ge\kern -0.1em V\!/}c}}\xspace}
\newcommand{\mevc}{\ensuremath{{\mathrm{\,Me\kern -0.1em V\!/}c}}\xspace}
\newcommand{\gevcc}{\ensuremath{{\mathrm{\,Ge\kern -0.1em V\!/}c^2}}\xspace}
\newcommand{\gevgevcccc}{\ensuremath{{\mathrm{\,Ge\kern -0.1em V^2\!/}c^4}}\xspace}
\newcommand{\mevcc}{\ensuremath{{\mathrm{\,Me\kern -0.1em V\!/}c^2}}\xspace}

\def\gsim{{~\raise.15em\hbox{$>$}\kern-.85em
          \lower.35em\hbox{$\sim$}~}\xspace}
\def\lsim{{~\raise.15em\hbox{$<$}\kern-.85em
          \lower.35em\hbox{$\sim$}~}\xspace}

\def\tell1  {TELL1\xspace}
\def\ukl1   {UKL1\xspace}



%% file: kmm-symbols-def.tex
\def\btosll               {\ensuremath{\decay{\bquark}{\squark\,\ellp\ellm}}\xspace}

\def\BbToKbandKstbll {\decay{\Bb}{\Kb^{(*)}\ell^+\ell^-}}
\def\BuToKmm      {\decay{\Bub}{\Km\mumu}}
\def\BuToKll      {\decay{\Bub}{\Km\ell^+\ell^-}}



%% file: title.tex
%
%
\begin{titlepage}

{\bf\boldmath\large
\begin{center}
  An empirical model to determine the hadronic resonance contributions
  to \BdbToKstmm transitions
\end{center}
}

\vspace*{2.0cm}

\begin{center}
T.~Blake$^1$, U.~Egede$^2$, P.~Owen$^3$, K.A.~Petridis$^4$,
G.~Pomery$^4$ \bigskip\\
{\it\footnotesize 
$ ^1$University of Warwick, Coventry, United Kingdom \\ \vspace{0.2cm}
$ ^2$Imperial College London, London, United Kingdom \\ \vspace{0.2cm}
$ ^3$Universit\"at Z\"urich, Z\"urich, Switzerland  \\ \vspace{0.2cm}
$ ^4$Univeristy of Bristol, Bristol, United Kingdom \\ \vspace{0.2cm}
}
\end{center}

\vspace{\fill}

\begin{abstract}
  \noindent
  A method for analysing the hadronic resonance contributions in
  \BdbToKstmm decays is presented. This method uses an empirical model
  that relies on measurements of the branching fractions and
  polarisation amplitudes of final states involving $J^{PC}=1^{--}$
  resonances, relative to the short-distance component, across the
  full dimuon mass spectrum of \BdbToKstmm transitions. The model is
  in good agreement with existing calculations of hadronic non-local
  effects.  The effect of this contribution to the angular observables
  is presented and it is demonstrated how the narrow resonances in the
  $\qsq$ spectrum provide a dramatic enhancement to \CP-violating
  effects in the short-distance amplitude.  Finally, a study of the
  hadronic resonance effects on lepton universality ratios,
  $R_{K^{(*)}}$, in the presence of new physics is presented.
  \end{abstract}

\vspace*{2.0cm}
\vspace{\fill}

\end{titlepage}




%% file: introduction.tex
\section{Introduction}
\label{sec:Introduction}

Decays with a \btosll transition receive contributions predominantly from loop-level, flavour changing neutral current transitions. These transitions are mediated by heavy (short-distance) particles and are suppressed in the Standard Model (SM).
Over the last few years, discrepancies have emerged when comparing measurements of the properties of \btosll decays to SM predictions~\cite{CMS:2017ivg,ATLAS:2017dlm,Wehle:2016yoi,LHCb-PAPER-2016-012,LHCb-PAPER-2016-051,LHCb-PAPER-2015-023,Khachatryan:2015isa,Lees:2015ymt,LHCb-PAPER-2014-007,LHCb-PAPER-2014-006}. Global analyses of these decays imply that there might be a new vector current which is destructively interfering with the SM contribution to the \btosll decay, producing inconsistency with the SM at the 4-5$\sigma$~\cite{Capdevila:2017bsm,Altmannshofer:2017fio,Hurth:2016fbr,Altmannshofer:2014rta,Straub:2015ica,Descotes-Genon:2015uva,Beaujean:2013soa}.

In this paper, the possibility that hadronic resonances are interfering with the short-distance amplitude and mimicking physics beyond the SM is considered. 
This is because in addition to the short-distance contribution to \btosll decays, the same final state can be obtained through non-local $b\to s q\overline{q}$ transitions, where $q\overline{q}$ denotes a quark-anti-quark pair. An example of such a decay is the decay \BdbToJPsiKst, where the \jpsi meson decays into two leptons.\footnote{Inclusion of charge conjugate processes is implied throughout this paper unless otherwise noted.} As the decay rate of this process is two orders of magnitude larger than its short-distance counterpart, sizeable interference effects are possible far from the \jpsi mass.

The approach presented in this paper models the hadronic contributions originating from charm and light quark resonances as Breit--Wigner amplitudes. This approach is inspired by Refs.~\cite{Lyon:2014hpa,Brass:2016efg} and is used to describe the hadronic resonances across the full dimuon mass spectrum of \BdToKstmm decays. The LHCb collaboration performed a measurement of the interference between the non-local and short-distance components of \BuToKmm decays by modelling the hadronic resonance contributions as Breit--Wigner amplitudes~\cite{LHCb-PAPER-2016-045}. The level of interference was found to be small and the measurement of the short-distance component was found to be compatible with that of previous interpretations.

These non-local contributions are difficult to calculate and to date there is no consensus as to whether the deviations seen in global analyses can be explained by the these intermediate hadronic contributions, or by physics beyond the SM. Differentiating between these two hypotheses is of prime importance for confirming the existence and subsequently characterising phenomena not predicted by the SM. More detailed discussions on this point can be found in Refs.~\cite{Bobeth:2017vxj,Capdevila:2017ert,Jager:2017gal,Khodjamirian:2017fxg,Brass:2016efg,Ciuchini:2015qxb,Lyon:2014hpa,Jager:2012uw,Khodjamirian:2012rm,Khodjamirian:2010vf}.

Due to the more complex amplitude structure of the decay, for each resonant final state there are three relative phases and magnitudes that need to be determined instead of one in the case of the \BuToKmm decay. Existing measurements of the branching fractions of \BdbToJPsiKst and \mbox{\BdbToPsiTwosKst} decays, together with measurements of their polarisation amplitudes~\cite{LHCb-PAPER-2013-059,Chilikin:2013tch,Aubert:2007hz,Chilikin:2014bkk} can be used to assess the impact of these decays to the observables of the \BdbToKstmm process, up to a single overall phase per resonance that needs to be determined through a simultaneous fit to both the short-distance and non-local components in the $\Kstarzb\mu^+\mu^-$ final state. In the absence of such a measurement, scanning over all possible values for the global phase for each resonant final state, results in a prediction of the range of hadronic effects that can be compared to more formal calculations. The angular distribution of the decay \BdbToKstmm is sensitive to the strong-phases of non-local contributions, particularly through the observables $S_{7}$ and $S_{9}$. This sensitivity allows for a data-driven extraction of the non-local parameters of the proposed model.

The level of \CP violation in decays such as \BdbToKstmm  depends on weak- and strong-phase differences with interfering processes, such as \BdbToJPsiKst. Therefore, a model for the strong phases of the non-local contributions to \BdToKstmm transitions, offers new insight on both the kinematic regions where \CP violation might be enhanced, as well as what the level of enhancement could be.

An increasingly large part of the discrepancy in \btosll transitions is being driven by tests of lepton universality in \BbToKbandKstbll decays~\cite{LHCb-PAPER-2017-013,Wehle:2016yoi,LHCb-PAPER-2014-024}. These deviations cannot be explained by hadronic effects (the \jpsi meson, for example, decays equally often to electrons and muons). Although a significant deviation from lepton-universality would be a clear indication of physics beyond the SM, the precise characterisation of the new physics model still depends on the treatment of hadronic contributions. The angular distribution of \BdbToKstll decays is critical in order to both determine the size of the new physics contribution, as well as to distinguish between models with left- or right-handed currents giving rise to new vector and axial-vector couplings.

This paper is organised as follows: Section~\ref{sec:model} describes the model of the non-local contributions as well as the experimental inputs; Section~\ref{sec:comparison} presents the comparison of the model to existing calculations; Section~\ref{sec:observables} shows how current model uncertainties
impact both \CP-averaged and \CP-violating observables of \BdbToKstmm decays, as well as the expected precision of the \BdbToKstmm observables using the data that is expected from the \lhcb experiment by the end of Run 2 of the LHC; finally in Section~\ref{sec:lnutests} there is a discussion of the impact of the non-local contributions in \BdbToKstll and \BuToKll transitions in the presence of lepton-universality violating physics.

%% file: method.tex
\section{The model}
\label{sec:model}

The differential decay rate of \BdbToKstmm transitions, where the \Kstarzb is a P-wave state and ignoring scalar or timelike contributions to the dimuon system, depends on eight independent observables~\cite{ref:egede:2008}. Each of these observables is made up of bilinear combinations of six complex amplitudes representing the three polarisation states of the \Kstarzb for both the left- and right-handed chirality of the dilepton system. The expression for the differential decay rate in terms of the angular observables  and their subsequent definition in terms of amplitudes, can be found in Ref.~\cite{Altmannshofer:2008dz}. The decay amplitudes are written in terms of the complex valued Wilson Coefficients $C_{7}$, $C_{9}$ and $C_{10}$, encoding short distance effects, and the $q^2$ dependent form-factors , \mbox{$F_{i}(q^2)=(V,A_{1},A_{12},T_{1},T_{2}, T_{23})$} given in Ref.~\cite{Straub:2015ica}, that express the $B\to K^{*}$ matrix elements of the operators involved in these decays. The coefficient $C_9$ corresponds to the coupling strength of the vector current operator, $C_{10}$ to the axial-vector current operator and $C_7$ to the electromagnetic dipole operator. A detailed review of these decays, including the operator definitions and the numerical values of the Wilson Coefficients in the SM, can be found in Ref~\cite{Altmannshofer:2008dz}. The decay amplitudes in the transversity basis and assuming a narrow $K^{*0}$ can be written as

\begin{dmath}
\label{eqn:ampstr1}
\mathcal{A}_{0}^{\rm L,R}(q^{2}) = -8N\frac{m_{B} m_{K^*}}{\sqrt{q^2}}\left\{(C_9\mp C_{10})A_{12}(q^2)+\frac{m_{b}}{m_{B}+m_{K^*}}C_{7}T_{23}(q^2)+\mathcal{G}_{0}(q^2)\right\},
\end{dmath}

\begin{dmath}
\label{eqn:ampstr2}
\mathcal{A}_{\parallel}^{\rm L,R}(q^{2}) = -N\sqrt{2}(m_{B}^{2}-m_{K^{*}}^{2})\left\{(C_9\mp C_{10})\frac{A_{1}(q^2)}{m_B-m_{K^*}}+\frac{2m_b}{q^2}C_{7}T_{2}(q^2)+\mathcal{G}_{\parallel}(q^2)\right\},
\end{dmath}

\begin{dmath}
\label{eqn:ampstr3}
\mathcal{A}_{\perp}^{\rm L,R}(q^{2}) = N\sqrt{2\lambda}\left\{(C_9\mp C_{10})\frac{V(q^2)}{m_B+m_{K^*}}+\frac{2m_b}{q^2}C_{7}T_{1}(q^2)+\mathcal{G}_{\perp}(q^2)\right\},
\end{dmath}
where $m_B$, $m_{K^*}$ and $m_\ell$ are the masses of the $B$-meson, $K^*$-meson, and lepton respectively, $q^2$ denotes the mass of the dimuon system squared, $\lambda=m_{B}^{4}+m_{K^*}^{4}+q^4-2(m_{B}^{2}m_{K^*}^{2}+m_{K^*}^{2}q^{2}+m_{B}^{2}q^{2})$, $\beta_\ell=\sqrt{1-4m_{\ell}^{2}/q^{2}}$ and 
\begin{equation}
N=V_{tb}V_{ts}^{*}\sqrt{\frac{G_{F}^{2}\alpha^{2}}{3\times2^{10}\pi^5m_{B}^{3}}q^2\lambda^{1/2}\beta_{\mu}}\, .
\end{equation}
\noindent In the above expressions, and for the remainder of this analysis, contributions from right handed Wilson Coefficients have been omitted. These are numerically small or zero in the SM and are not currently favoured by global analyses of \btosll processes. Following Ref.~\cite{Straub:2015ica}, the form factors are written  

\begin{equation}
 \label{eq:formfactors}
F^{i}(q^2)=\frac{1}{1-q^2/m_{R_i}^2}\sum_{k=0}^{2}\alpha_{k}^{i}[z(q^2)-z(0)]^k,
\end{equation}
where the $z$ function is given by
 \begin{equation}
 \label{eq:zexpansion}
 z(t) = \frac{\sqrt{t_{+} -t}- \sqrt{t_{+}-t_{0}}}{\sqrt{t_{+} -t}+ \sqrt{t_{+}-t_{0}}} \, ,
 \end{equation}
with $t_{\pm} = (m_{B} \pm m_{K^*})^{2}$ and $t_{0} = t_{+} (1- \sqrt{1- t_{-}/t_{+}})$.
The parameters $m_{R_i}$ are taken from Ref.~\cite{Straub:2015ica} and the coefficients $\alpha_{k}^{i}$ including their correlations are taken from a combined fit to light-cone sum rule calculations and Lattice QCD results given in Refs.~\cite{Straub:2015ica,Horgan:2013hoa}. 

The functions $\mathcal{G}_{\lambda}(q^2)$ describe the non-local hadronic contributions to the $\BdToKstmm$ amplitudes and are given by a simplistic empirical parametrisation inspired by the procedure of Refs~\cite{LHCb-PAPER-2016-045,Lyon:2014hpa, Brass:2016efg}. In particular

\begin{dmath}
\label{eqn:h1}
\mathcal{G}_{0}=\frac{m_b}{m_{B}+m_{K^{*}}}T_{23}(q^2)\zeta^0e^{i\omega^0}+A_{12}(q^2)\sum_j \eta_{j}^{0}e^{i\theta_{j}^{0}}A_{j}^{\rm res}(q^2),
\end{dmath}

\begin{dmath}
\label{eqn:h2}
\mathcal{G}_{\parallel}=\frac{2m_b}{q^2}T_{2}(q^2)\zeta^\parallel e^{i\omega^\parallel}+\frac{A_{1}(q^2)}{m_B-m_{K^*}}\sum_j \eta_{j}^{\parallel}e^{i\theta_{j}^{\parallel}}A_{j}^{\rm res}(q^2),
\end{dmath}

\begin{dmath}
\label{eqn:h3}
\mathcal{G}_{\perp}=\frac{2m_b}{q^2}T_{1}(q^2)\zeta^\perp e^{i\omega^\perp}+\frac{V(q^2)}{m_B+m_{K^*}}\sum_j \eta_{j}^{\perp}e^{i\theta_{j}^{\perp}}A_{j}^{\rm res}(q^2),
\end{dmath}
\noindent where the sum in the above expressions represents the coherent sum of vector meson resonant amplitudes with
$\eta_j^{\lambda}$ and $\theta_j^{\lambda}$ the magnitude and phase of each resonant amplitude relative to $C_9$. The exact normalisation of the $\eta_{j}^{\lambda}$ parameters is shown in Appendix~\ref{app:norm} and is chosen 
such the integral of the sum of the squared magnitudes of the amplitude of a given amplitude produce the correct experimental branching fraction. Similarly, the parameters $\zeta^{\lambda}$ and $\omega^{\lambda}$ need to be determined from experimental measurements. In this analysis, the central values of these parameters are set to zero, unless otherwise specified.

The $q^2$ dependence of each resonant amplitude is given by $A_j^{\rm res}(q^2)$. As indicated from the analysis of the dimuon mass spectrum in \BuToKmm decays from Ref.~\cite{LHCb-PAPER-2016-045}, the resonances considered in this analysis are the $\rho^0$, $\phi$, $\jpsi$, $\psi(2S)$, $\psi(3770)$, $\psi(4040)$ and $\psi(4160)$. Contributions from light-quark resonances are expected to be either CKM- or loop-suppressed compared to final states occurring through charmonium resonances. 
As experiments accumulate more data, additional broad light-quark states will start becoming statistically significant and can be easily incorporated in the model.
For simplicity, $A_j^{\rm res}$ is modelled by a relativistic Breit--Wigner function given by

\begin{equation}
A_j^{\rm res}(q^2) = \frac{m_{{\rm res}\,j}\Gamma_{{\rm res}\,j}}{(m_{{\rm res}\,j}^{2}-q^2)-i m_{{\rm res}\,j}\Gamma_{j}(q^2)},
\end{equation}
where $m_{{\rm res}\,j}$ and $\Gamma_{{\rm res}\,j}$ are the pole mass and natural width of the $j^{\rm th}$ resonance and their values are taken from Ref.~\cite{PDG2016}. The running width $\Gamma_{j}(q^2)$ is given by

\begin{equation}
\Gamma_{j}(q^2)=\frac{p}{p_{{\rm res}\, j}}\frac{m_{{\rm res}\,j}}{q}\Gamma_{{\rm res}\,j},
\end{equation}

\noindent where $p$ is the momentum of the muons in the rest frame of the dimuon system evaluated at $q$, and $p_{{\rm res}\, j}$ is the momentum evaluated at the mass of the resonance. 

This isobar approach, although not rigorous, it provides a model for the strong phase variation of the amplitude across the full $q^2$ spectrum.
This variation can result in sizeable effects even far from the pole of the resonances as discussed in Sec.~\ref{sec:observables}. 

It is customary that for each helicity amplitude, the expressions of the non-local components $\mathcal{G}_{\lambda}$ are recast as shifts to the Wilson coefficient $C_9$, referred to as $\Delta C_{9\,\,\lambda}^{\rm total}$. This convention is particularly useful for comparisons with formal predictions of the non-local contributions. 

Measurements of $\Bdb\to V \Kstarzb$ decays, where $V$ denotes any $J^{PC}=1^{--}$ state, are only sensitive to relative phases of the three transversity amplitudes.  Therefore, the convention used in previous measurements of these modes is such that phases $\theta_\parallel$ and $\theta_\perp$ are defined relative to $\theta_0$. Using this convention, the remaining phase difference of each resonant polarisation amplitude relative to the corresponding short-distance one, is given by $\theta_0$.

\subsection{Experimental input}

In order to assess the impact of the resonances appearing in the dimuon spectrum of \BdbToKstmm decays, knowledge of the resonance parameters $\eta_{j}$ and $\theta_{j}$ appearing in Eqs.~\ref{eqn:h1}--\ref{eqn:h3} is required. The amplitude analyses of \BdbToJPsiKst and \BdbToPsiTwosKst transitions performed by the LHCb, BaBar and Belle collaborations~\cite{aaij2013measurement,Chilikin:2013tch,Aubert:2007hz} constrain the relative phases and magnitudes of the transversity amplitudes of the resonant decay modes. Combined with the measured branching fractions of these decays by the Belle experiment~\cite{Chilikin:2013tch,Chilikin:2014bkk}, the parameters $\eta_{j}^{\parallel,\perp,0}$ and $\theta_{j}^{\parallel,\perp}$ are determined up to an overall phase, $\theta_{j}^{0}$, relative to the short-distance amplitude for the \BdbToKstmm decay. Similarly, the amplitude components of \BdbToPhiKst transitions have been determined up to an overall phase, through the amplitude analyses and branching fraction measurements given in Refs.~\cite{LHCb-PAPER-2014-005,Prim:2013nmy,Aubert:2008zza}. 

For the decay \BdbToRhoKst, the magnitude of the total decay amplitude is set using the world average branching fraction of this transition~\cite{PDG2016,Lees:2011dq,Kyeong:2009qx}. As no amplitude analysis of this mode has been performed, the relative phases and magnitudes of the transversity amplitudes are taken to be the same as those of the \BdbToPhiKst decay. As the overall contribution of the $\rho^0$ is expected to be small, this assumption will not impact the main conclusions of this study.

No measurements exist for final states involving the $\psi(3770)$, $\psi(4040)$ and $\psi(4160)$ resonances, denoted as $\Bdb\to V_{\psi} \Kstarzb$. To estimate the contributions of these final states, the relative phases and magnitudes of the transversity amplitudes are taken from the amplitude analysis of \BdbToJPsiKst decays. An approximate value of the branching fraction of each of the $\Bdb \to V_{\psi} \Kstarzb$ modes is obtained by scaling the measured branching fraction of the decay \BdbToPsiTwosKst, with $\psitwos\to\mup\mun$, by the known ratio of branching fractions between $B^+\to\psitwos K^+$ and $B^+\to V_{\psi} K^+$ decays, with $V_{\psi}\to\mup\mun$,  given in Ref.~\cite{LHCb-PAPER-2016-045}. The values used for the relative amplitudes and phases for each resonant contribution are summarised in Table~\ref{tab:res_vals}.

\begin{table}[t]
  \centering
    \begin{tabular}{l c c c}
      Mode & $(\eta_{j}^{\parallel},\theta_{j}^{\parallel} \,[rad])$ & $(\eta_{j}^{\perp},\theta_{j}^{\perp}\,[rad])$ & $\eta_{j}^{0}$ \\
      \hline
      $B^0\to\rho^{0}K^{*0}$       & $(1.5\phantom{\times10^{++0}},\phantom{+}2.6)$  & $(1.9\phantom{\times10^{++0}},\phantom{+}2.6)$ &  $5.1\times10^{-1}$ \\
      $B^0\to\phi K^{*0}$           & $(2.5\times10^{+1},\phantom{+}2.6)$       & $(3.2\times10^{+1},\phantom{+}2.6)$    &  $1.0\times10^{+1}$      \\
      $B^0\to\jpsi K^{*0}$           & $(4.9\times10^{+3},-2.9)$    & $(6.5\times10^{+3},\phantom{+}2.9)$     &  $7.1\times10^{+3}$    \\
      $B^0\to\psitwos K^{*0}$    & $(5.3\times10^{+2},-2.8)$   & $(8.1\times10^{+2},\phantom{+}2.8)$   &  $9.6\times10^{+2}$    \\
      $B^0\to\psi(3770) K^{*0}$ & $(9.3\times10^{-1},-2.9)$  & $(1.5\phantom{\times10^{++0}},\phantom{+}2.9)$ &  $1.7\phantom{\times10^{++0}}$  \\
      $B^0\to\psi(4040) K^{*0}$ & $(2.9\times10^{-1},-2.9)$  & $(5.6\times10^{-1},\phantom{+}2.9)$ &  $6.0\times10^{-1}$  \\
      $B^0\to\psi(4160) K^{*0}$ & $(8.3\times10^{-1},-2.9)$  & $(2.0\phantom{\times10^{++0}},\phantom{+}2.9)$ &  $1.8\phantom{\times10^{++0}}$\\
    \end{tabular}
    \caption{Summary of the input values used to model the non-local amplitude components $\mathcal{G}_{\lambda}$.
      The input values rely on measurements given in Refs.~\cite{LHCb-PAPER-2016-045,LHCb-PAPER-2013-059,Chilikin:2013tch,Aubert:2007hz,Chilikin:2014bkk,LHCb-PAPER-2014-005,Prim:2013nmy,Aubert:2008zza,Lees:2011dq,Kyeong:2009qx}. 
    The phases are measured relative to $\theta_{j}^{0}$. As the measurements are given for the decay of the \Bd meson,
      in order to convert to the decay of the \Bdb, the phase $\theta_{j}^{\perp}$ given in the table above must be shifted by $\pi$.}
    \label{tab:res_vals}
  \end{table}

%% file: model_comp.tex
\section{Model comparisons}
\label{sec:comparison}

The study presented in Ref.~\cite{Khodjamirian:2010vf} provides a prediction of the non-local charm loop contribution to \BdbToKstmm decays. It relies on QCD light-cone sum rule calculations of $B\to K^{*}$ matrix elements for $q^2\ll4m_{c}^{4}$ and extrapolated to larger $q^2$ through a hadronic dispersion relation. The extrapolation uses input from experimental measurements of the rate and amplitude structure of \BdbToJPsiKst and \BdbToPsiTwosKst decays. As this calculation does not account for the factorisable next-to-leading order corrections to the charm loop, all phases of the non-local relative to the short-distance amplitudes are set to zero. 

\begin{figure}[!!!t]
\centering
\includegraphics[width=0.48\textwidth]{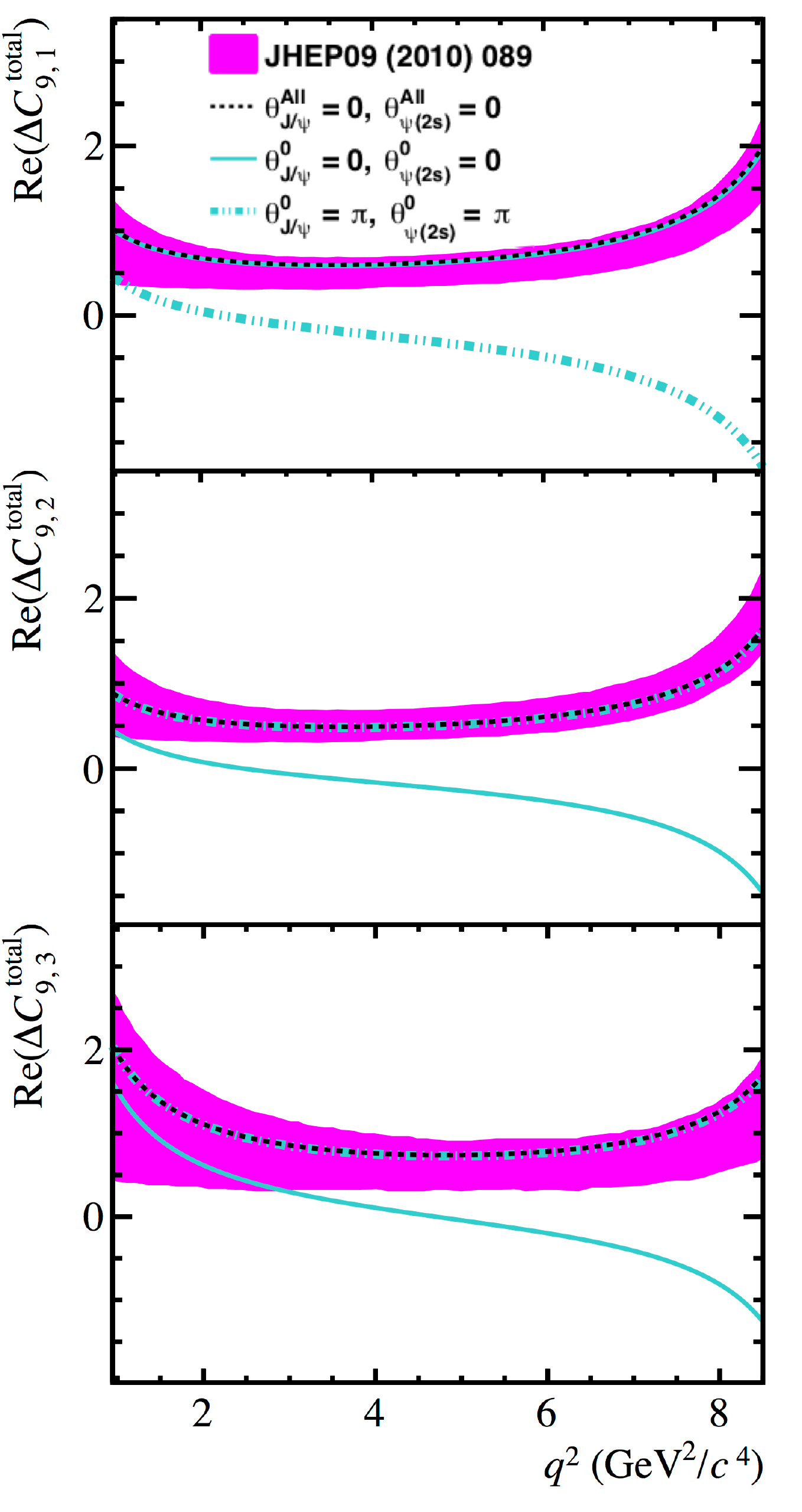}
\caption{The non-local contributions to the \BdToKstmm invariant amplitudes as a function of $q^2$. The prediction using the model discussed in Sec.~\ref{sec:model} is shown, where only the contributions from the \jpsi and \psitwos resonances are considered. The free phases $\theta^{0}_{\jpsi}$ and $\theta^{0}_{\psitwos}$ are both set to 0 (cyan solid line) or $\pi$ (cyan dashed-dotted line). The prediction where all phases of the \jpsi and \psitwos appearing in Eqs.~\ref{eqn:h1}--\ref{eqn:h3} are set to zero is also depicted (black solid line), alongside the prediction from Ref.~\cite{Khodjamirian:2010vf} (magenta band).    }\label{fig:dc9_kstmm_kmpw_comp}
\end{figure}

Figure~\ref{fig:dc9_kstmm_kmpw_comp} shows the parametrisation of the non-local contributions in the invariant amplitude basis of \BdbToKstmm decays given in Ref.~\cite{Khodjamirian:2010vf}. The relation of this amplitude basis to the helicity basis is also given in~Ref.~\cite{Khodjamirian:2010vf}. The predictions using the model described in Sec.~\ref{sec:model}, where only the contributions from the \jpsi and \psitwos resonances are considered, are shown for comparison. The free phases $\theta^{0}_{\jpsi}$ and $\theta^{0}_{\psitwos}$ appearing in Eqs.~\ref{eqn:h1}--\ref{eqn:h3} are both set to 0 or $\pi$. As a consistency check, the model presented in this paper is also shown, with the phases of all transversity amplitudes set to zero. The parameters $\zeta_{\lambda}$ and $\omega_{\lambda}$ also appearing in Eqs.~\ref{eqn:h1}--\ref{eqn:h3} are chosen such that they are broadly consistent with the values of  Ref.~\cite{Straub:2015ica} and the predictions of Ref.~\cite{Khodjamirian:2010vf}, with $\zeta_{\lambda}\sim0.08|C_{7}|$ and $\omega_{\lambda}=\pi$. Ignoring all phases of the transversity amplitudes of \BdbToJPsiKst and \BdbToPsiTwosKst decays, the model of $\Delta C_{9\,\,\lambda}^{\rm total}$ described in this analysis is consistent to that of Ref.~\cite{Khodjamirian:2010vf}. However, accounting for the measured relative phases in the resonant  decay amplitudes results in large differences between the two models. The level of disagreement depends on the value of the free phases $\theta^{0}_{\jpsi}$ and $\theta^{0}_{\psitwos}$. The effect of the non-local charm contributions in Ref.~\cite{Khodjamirian:2010vf} are known to move the central value of predictions of angular observables such as $P_{5}'$ further away from experimental measurements~\cite{Descotes-Genon:2015uva}. However, this effect is only true due to the fact that the analysis of Ref.~\cite{Khodjamirian:2010vf}  did not account for the phases of the resonant amplitudes. An assessment of the impact of the phases on the angular observables is discussed in Sec.~\ref{sec:observables}

\begin{figure}[!htb]
\centering
\includegraphics[width=1\textwidth]{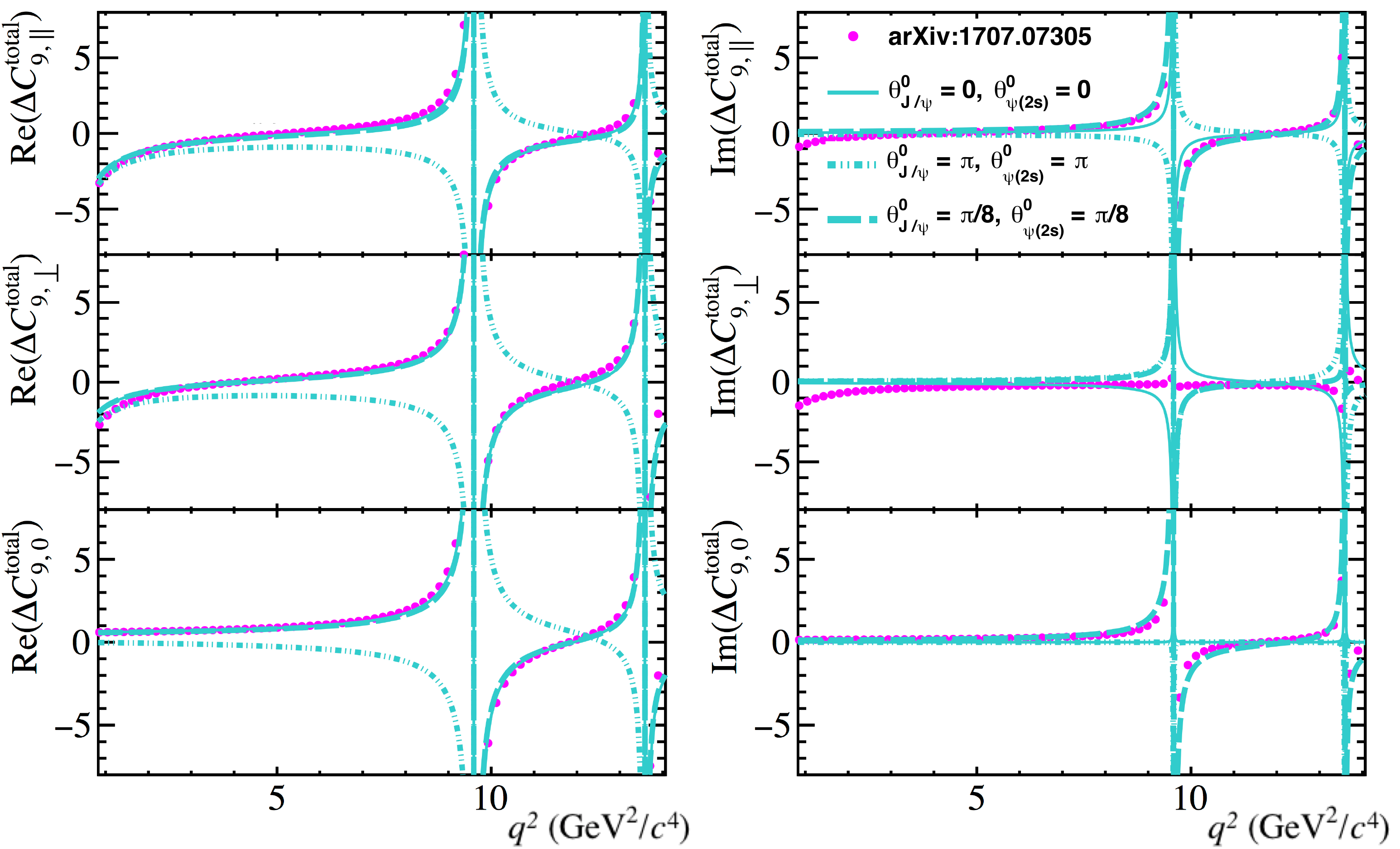}
\caption{The non-local contributions to the \BdbToKstmm transversity amplitudes as a function of $q^2$. The real (left) and imaginary (right) components are shown separately. The prediction from Ref.~\cite{Bobeth:2017vxj} is shown (magenta points). Predictions using the model discussed in Sec.~\ref{sec:model}, where only the contributions from the \jpsi and \psitwos resonances are considered, are  overlaid for different choices of the phases $\theta^{0}_{\jpsi}$ and $\theta^{0}_{\psitwos}$ (cyan lines). See text for further details.}\label{fig:dc9_kstmm_vDBJC17_comp}
\end{figure}

Building on the ideas of Ref.~\cite{Khodjamirian:2010vf}, a recent analysis presented in Ref.~\cite{Bobeth:2017vxj} provides a prediction of the non-local charm contribution that is valid up to a $q^2\leq m_{ \psitwos}^{2}$. This prediction also makes use of experimental measurements of \BdbToJPsiKst and \BdbToPsiTwosKst decays. In contrast to Ref.~\cite{Khodjamirian:2010vf}, the calculations of the non-local contributions are performed at  $q^2<0$ to next-to-leading order in $\alpha_{s}$. The $q^2$ parametrisation is given by a $z$-expansion truncated after the second order as in Eq.~(\ref{eq:formfactors}). Figure~\ref{fig:dc9_kstmm_vDBJC17_comp} shows both the real and imaginary parts of the non-local contributions to \BdbToKstmm decays presented in Ref.~\cite{Bobeth:2017vxj}. As the correlations between the $z$-expansion parameters are not provided, only the central values of the predictions are shown. The phase convention used in Ref.~\cite{Bobeth:2017vxj} is such that the transversity amplitudes of the \BdbToJPsiKst and \BdbToPsiTwosKst decays are related to those presented in this study through $\eta^{\parallel}_{j}\to -\eta^{\parallel}_{j}$. 
The model described in Sec.~\ref{sec:model}, where only the contributions from the \jpsi and \psitwos resonances are considered, is in qualitative agreement with that of Ref.~\cite{Bobeth:2017vxj} for the following parameter choice: $\theta^{0}_{\jpsi}=\pi/8$, $\theta^{0}_{\psitwos}=\pi/8$, $\zeta_{\lambda}\sim15\%|C_{7}|$ and $\omega_{\lambda}=\pi$.  The small level of disagreement observed in the imaginary part of the amplitudes at low $q^2$ is due to the choice of setting $\omega_{\lambda}=\pi$, with smaller values giving a better agreement. 

To conclude, the simplistic model of the non-local contributions to \BdbToKstmm decays presented in this paper is in good agreement with existing models, provided appropriate choices of $\theta^{0}_{\jpsi}$, $\theta^{0}_{\psitwos}$, $\omega_{\lambda}$ and $\zeta_{\lambda}$. For the latter, a larger value is  required to match the predictions of Ref.~\cite{Bobeth:2017vxj}, compared to Ref.~\cite{Khodjamirian:2010vf}. The expressions of $\mathcal{G}_{\lambda}(q^2)$ have sufficient freedom to capture the $q^2$ dependence of formal theory predictions in the $q^2$ range $1<q^2<m_{\psitwos}^{2}$. In addition, in contrast to current predictions, the model of  $\Delta C_{9\,\,\lambda}^{\rm total}(q^2)$ can naturally accommodate hadronic contributions from $J^{PC}=1^{--}$ states composed of light quarks such as the $\phi$ and $\rho^0$, as well as resonances appearing in the region $q^2>4m_{D}^{2}$, where $m_{D}$ denotes the mass of the $D$-meson. This is due to the use of Breit--Wigner functions to approximate the resonant contributions, that experiments can easily adopt. 

%% file: observables.tex
\section{Effect on {\boldmath \BdbToKstmm} angular observables}
\label{sec:observables}
\begin{figure}
  \centering
  \includegraphics[width=1\textwidth]{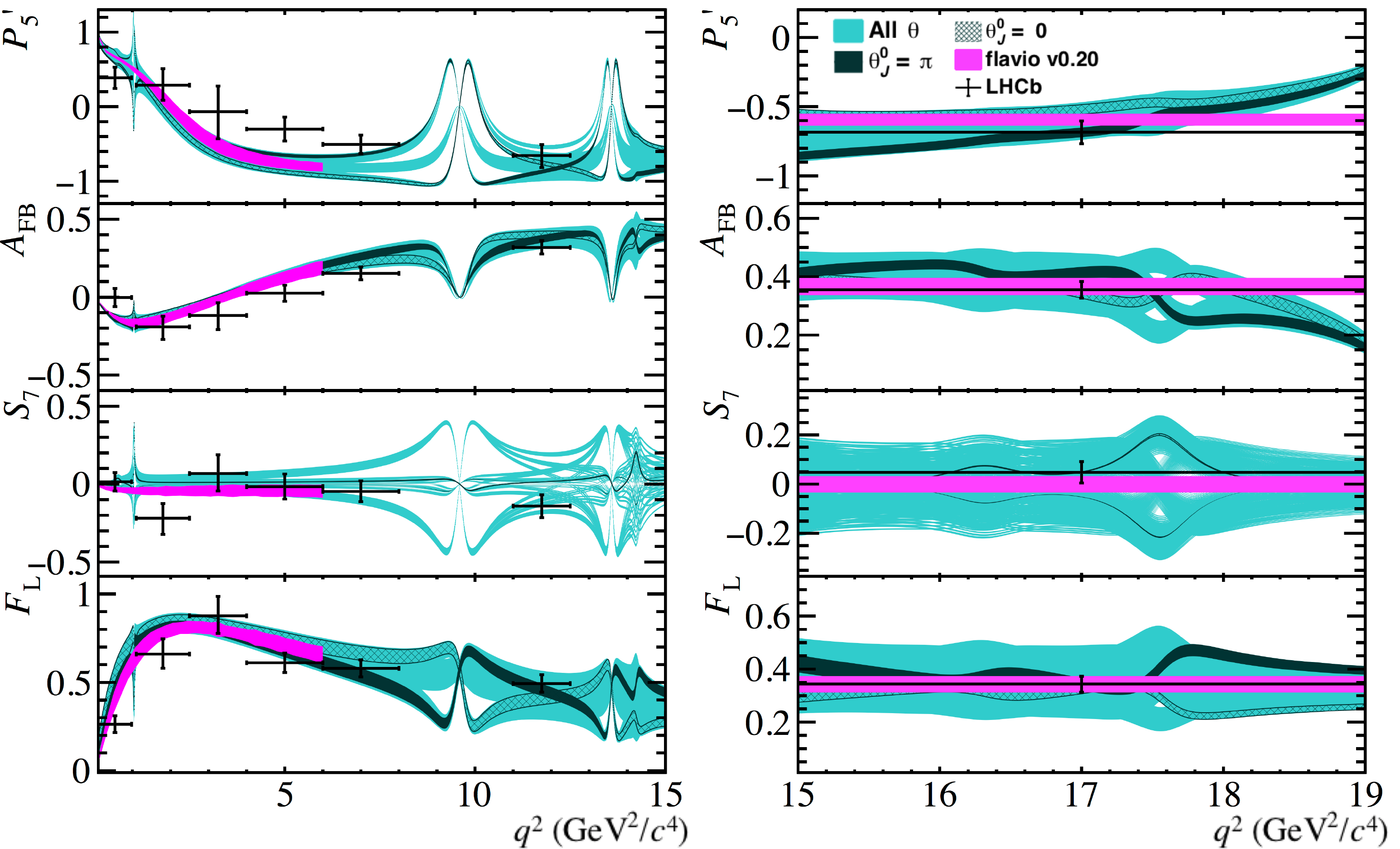}
  \caption{Distributions of the angular observables $P_{5}^{\prime}$, $A_{\rm FB}$ $S_{7}$, and $F_{L}$ as a function of $q^2$ for regions below (left) and above (right) the open charm threshold (cyan). Specific choices are highlighted for $\theta_{j}^{0}=0$ (hatched band) and $\theta_{j}^{0}=\pi$ (dark band). The measured values of the observables from Ref.~\cite{LHCb-PAPER-2015-051} are also shown (black points). The theoretical predictions (magenta band) using \texttt{flavio}~\cite{ref:flavio} are shown for comparison.
  }\label{fig:p5p}
\end{figure}

Using the model of $\Delta C_{9\,\,\lambda}^{\rm total}$ described in Sec.~\ref{sec:model}, the effect of the hadronic resonance contributions on the angular observables of \BdbToKstmm decays can be estimated. Figure~\ref{fig:p5p} shows the distribution of the angular observables $P_{5}^{\prime}$, $A_{\rm FB}$, $S_7$ and $F_{L}$~\cite{DescotesGenon:2012zf,Kruger:1999xa} in the SM. The observable $S_7$ exhibits a particularly large dependence on the strong phases, demonstrating that measurements of the angular distribution of \BdbToKstmm decays can be used to determine the phases of the hadronic resonances. Therefore, this observable can be used to separate short-distance from the non-local contributions, as only the non-local part has a strong-phase difference. The remaining \CP-averaged observables can be found in Appendix~\ref{app:cp_observables}. Definitions of these observables can be found for instance in Ref.~\cite{Descotes-Genon:2013vna}. As the phase $\theta_{j}^{0}$ of all the resonant final states appearing in Table~\ref{tab:res_vals} are unknown, all possible variations of phases $\theta_{j}^{0}$ are considered. The uncertainties arising from the combined light-cone sum rules and lattice QCD calculations of $B\to K^{*}$ form factors are accounted for using the covariance matrix provided in Ref.~\cite{Straub:2015ica}. The predictions of these observables using \texttt{flavio}~\cite{ref:flavio} are also shown for comparison. The lack of knowledge of the phase $\theta_{j}^{0}$ results in a large uncertainty for the prediction of $P_{5}'$, diluting the sensitivity of this observable to the effects of physics beyond the SM. However, for the choice of $\theta_{j}^{0}$ that results in a non-local charm contribution that is compatible with the latest prediction presented in Ref.~\cite{Bobeth:2017vxj}  and is shown in Fig.~\ref{fig:dc9_kstmm_vDBJC17_comp}), the tension of the prediction with the measured value of $P_{5}'$ cannot be explained solely through hadronic effects.


\subsection{Sensitivity to {\boldmath \CP} violation}
\label{sec:cp_asym}
The model of the hadronic resonance contributions to \BdbToKstmm decays described in this paper provides
a prediction for the strong phase differences involved in these transitions. 
Direct \CP violation will arise when there are interfering amplitudes that have different weak phases as well as different strong phases, as discussed within the context of \BuToKmm and \BuToPimm decays in Refs.~\cite{Khodjamirian:2017fxg,Hambrock:2015wka}. Therefore, it is interesting to study the effect that potential weak phases beyond the SM have on angular observables such as the direct \CP asymmetry $A_{\CP}$, defined as
\begin{equation}
\displaystyle A_{\CP}=\dfrac{\dfrac{d\Gamma(B\to
   K^{*}\mu^+\mu^-)}{dq^2}-\dfrac{d\overline{\Gamma}(B\to
    K^{*}\mu^+\mu^-)}{dq^2}}{\dfrac{d\Gamma(B\to
    K^{*}\mu^+\mu^-)}{dq^2}+\dfrac{d\overline{\Gamma}(B\to
    K^{*}\mu^+\mu^-)}{dq^2}},
\end{equation}
where $\Gamma$ and $\overline{\Gamma}$ correspond to the partial widths of the decays \BdbToKstmm and \BdToKstmm respectively, as well as the so-called \CP-odd angular observables $A_{i}$, defined for instance in Ref.~\cite{Altmannshofer:2008dz}. The method is similar to what is discusssed in Ref.~\cite{Fajfer:2012nr} for semileptonic charm decays.

\begin{figure}[!t]
  \centering
  \includegraphics[width=0.49\textwidth]{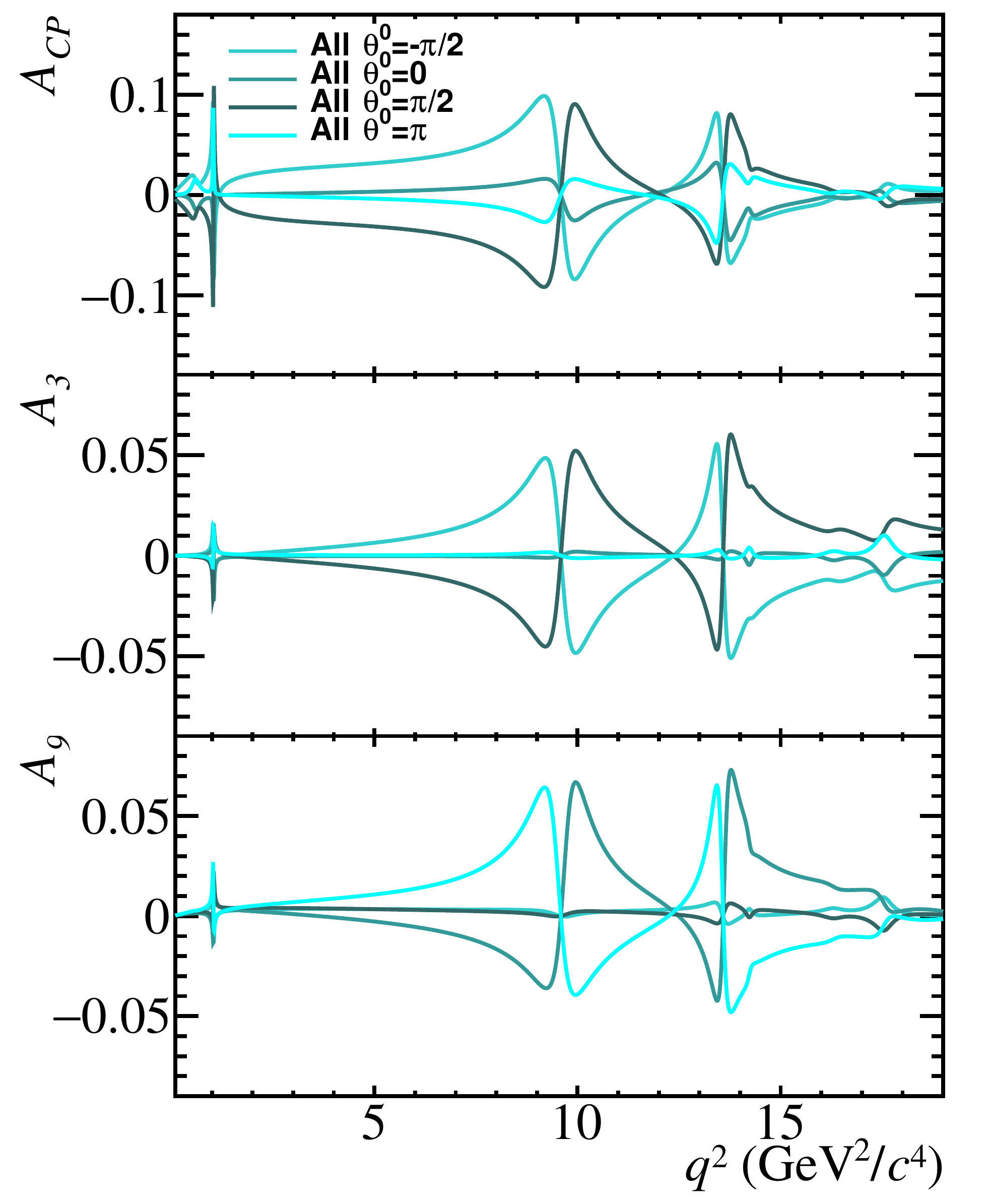}
  \includegraphics[width=0.49\textwidth]{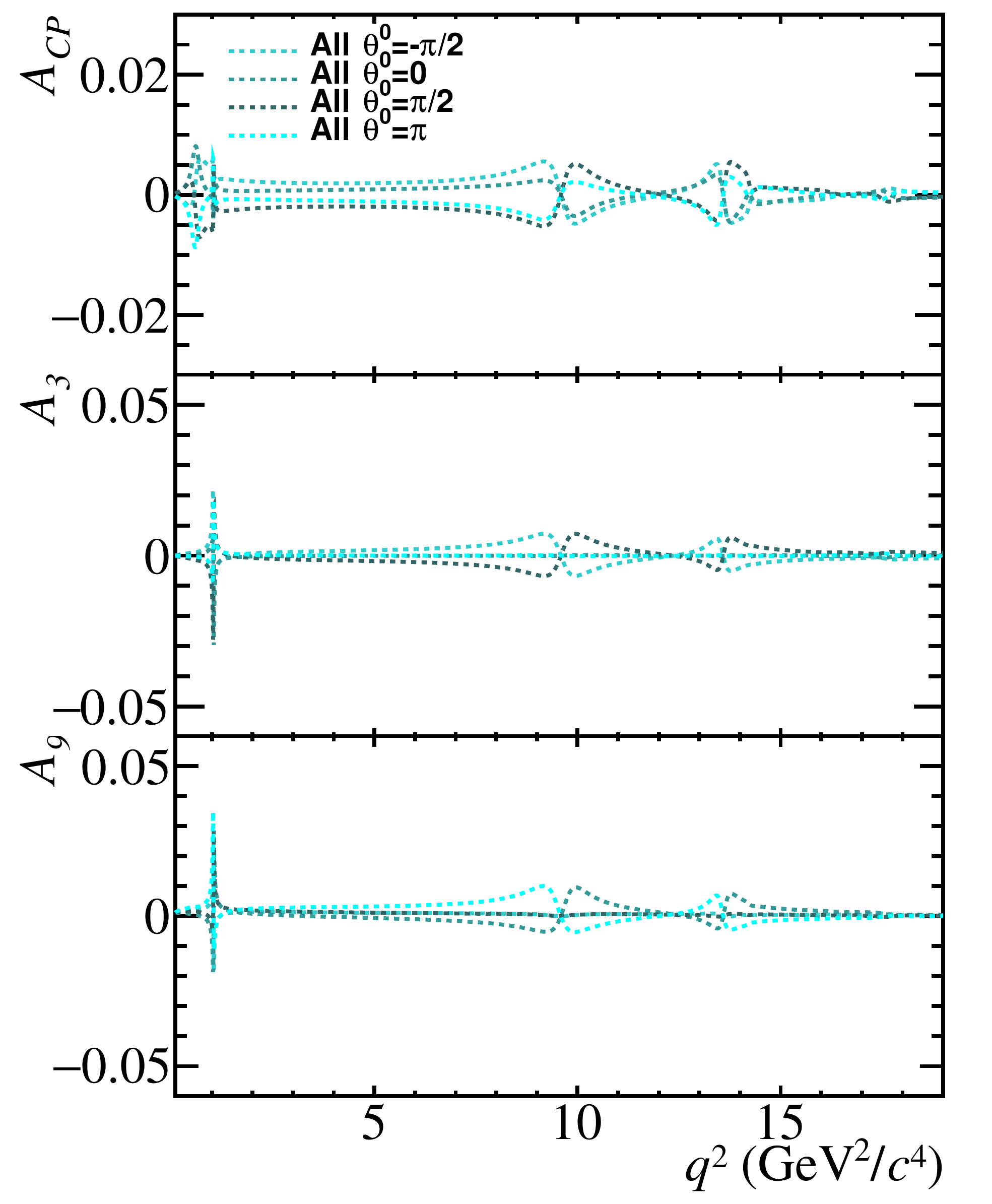}
  \caption{  Distribution of observables $A_{\CP}$, $A_{3}$ and $A_{9}$ as a function of $q^2$, for $\theta_j^{0}$ of all resonances set to $-\pi/2$, $0$, $\pi/2$ and $\pi$. Two new physics models are considered, one with $C_{9}^{NP}=-1.0-1.0i$ (left), and one with $C_{7}^{NP}=-0.03i$, $C_{9}^{NP}=-1.0$ (right). }\label{fig:acp_figs}
\end{figure}

Figure~\ref{fig:acp_figs} shows the observables $A_{\CP}$, $A_{3}$ and $A_{9}$ for 
$\theta_j^{0}$ of all resonances set to $-\pi/2$, $0$, $\pi/2$ and $\pi$. To illustrate the effect that the model of the strong phase differences have in the presence of new weak phases, two new physics models are considered which are compatible with existing experimental constraints. One with $C_{9}^{\rm{NP}}=-1.0-1.0i$, and one with both $C_{7}^{\rm{NP}}=-0.03i$ and $C_{9}^{\rm{NP}}=-1.0$~\cite{Alok:2017jgr,Paul:2016urs}. The notation $C_{7,9}^{\rm{NP}}$ denotes the new physics contribution to the corresponding Wilson Coefficient. In both these models, all other Wilson Coefficients are set to their SM values. It is clear that the non-local contribution enhances \CP-violating effects in these decays, with the level of this enhancement depending on the value of the unknown phase $\theta_j^{0}$. As it can be seen, there is a huge effect in the vicinity of the resonances, thus giving sensitivity to an imaginary component of $C_9$ in a way which have not been considered before. The only other viable way to gain sensitivity would be through a time dependent analysis of the $\decay{\Bs}{\jpsi\phi}$ or the \BdToKstmm with the $\Kstarz$ decaying to the \CP eigenstate $\KS\piz$.  In contrast, \CP-violating effects arising through a weak phases appearing in the Wilson coefficient $C_7$, are best constrained from measurements of $B\to K^*\gamma$ decays~\cite{Paul:2016urs}.

\subsection{Expected experimental precision}
\label{sec:exp_prec}
The experimental sensitivity to the phases between the short-range and hadronic resonance contributions to \BdbToKstmm is determined using $\mathcal{O}(10^{6})$ simulated decays that include contributions from both short-distance and non-local components. The size of this sample corresponds to the approximate number of decays expected\footnote{The yield of both short-distance and non-local $\BdbToKstmm$ decays is calculated by scaling the number of $\BdbToJPsiKst$ and short-distance $\BdbToKstmm$ decays given in Ref.~\cite{LHCb-PAPER-2015-051} by a factor of 4.} to be collected by the LHCb experiment by the end of Run2 of the LHC~\cite{LHCb-PAPER-2015-051}. The decays are generated with the parameters $\theta_{j}^{0}$, $\zeta_\lambda$ and $\omega_\lambda$ set to zero. The S-wave contribution to \BdbToKstmm decays is accounted for using the angular terms and amplitude expressions as a function of the invariant mass of the $K\pi$ system given in Refs.~\cite{ref:wang_swave,ref:wang_kappa}.
In addition to the S-wave component for the short-range amplitude, S-wave components are introduced with an amplitude and phase ($\eta_{j}^{S}$, $\theta_{j}^{S}$), for the \jpsi and the \psitwos resonances, based on the measurements given in Refs.~\cite{LHCb-PAPER-2013-059,Chilikin:2013tch}. The overall effect of the S-wave contribution to the remaining resonances is considered to be negligible and is therefore ignored. In this study, all Wilson Coefficients are assumed to be real.

In order to ascertain the statistical precision on the non-local contribution, the detector resolution in $q^2$ needs to be accounted for by smearing the $q^2$ spectrum of the simulated events. For simplicity, a Gaussian resolution function is used with a width based on the RMS value of the dimuon mass resolution provided in Ref.~\cite{LHCb-PAPER-2016-045}, and converted into a resolution in $q^2$. As the resolution in the helicity angles are far better than the variations in the angular distributions, any resolution effect in angles can be ignored;
the sharp shape of the $\phi$, \jpsi and \psitwos resonances mean that a similar argument is not valid for the $q^2$ distribution.

A four dimensional maximum likelihood fit is performed to the $q^2$, \ctl, \ctk and $\phi$ distributions of the \BdbToKstmm decays in this sample. Both the non-local parameters, including $\eta_{j}^{S}$ and $\theta_{j}^{S}$, as well as the Wilson Coefficients $C_9$ and $C_{10}$ are left to vary in the fit. The $B\to K^*$ form factor parameters however are fixed to their central values given in Ref.~\cite{Straub:2015ica}.  The resulting covariance matrix is used to ascertain the statistical precision on $\Delta C_{9\,\,\lambda}^{\rm total}$. Based on the assessment of the systematic uncertainties in Ref.~\cite{LHCb-PAPER-2016-045}, the dominant source of experimental uncertainty is expected to be statistical in nature. However, the presence of tetra-quark states appearing in \BdbToJPsiKPi and \BdbToPsiTwosKPi decays~\cite{Aaij:2014jqa,Chilikin:2013tch} will impact the determination of the non-local parameters. Although the effect is expected to be small, an accurate assessment of the effect is beyond the scope of this study.

\begin{figure}[!t]
  \centering
  \includegraphics[width=1.\textwidth]{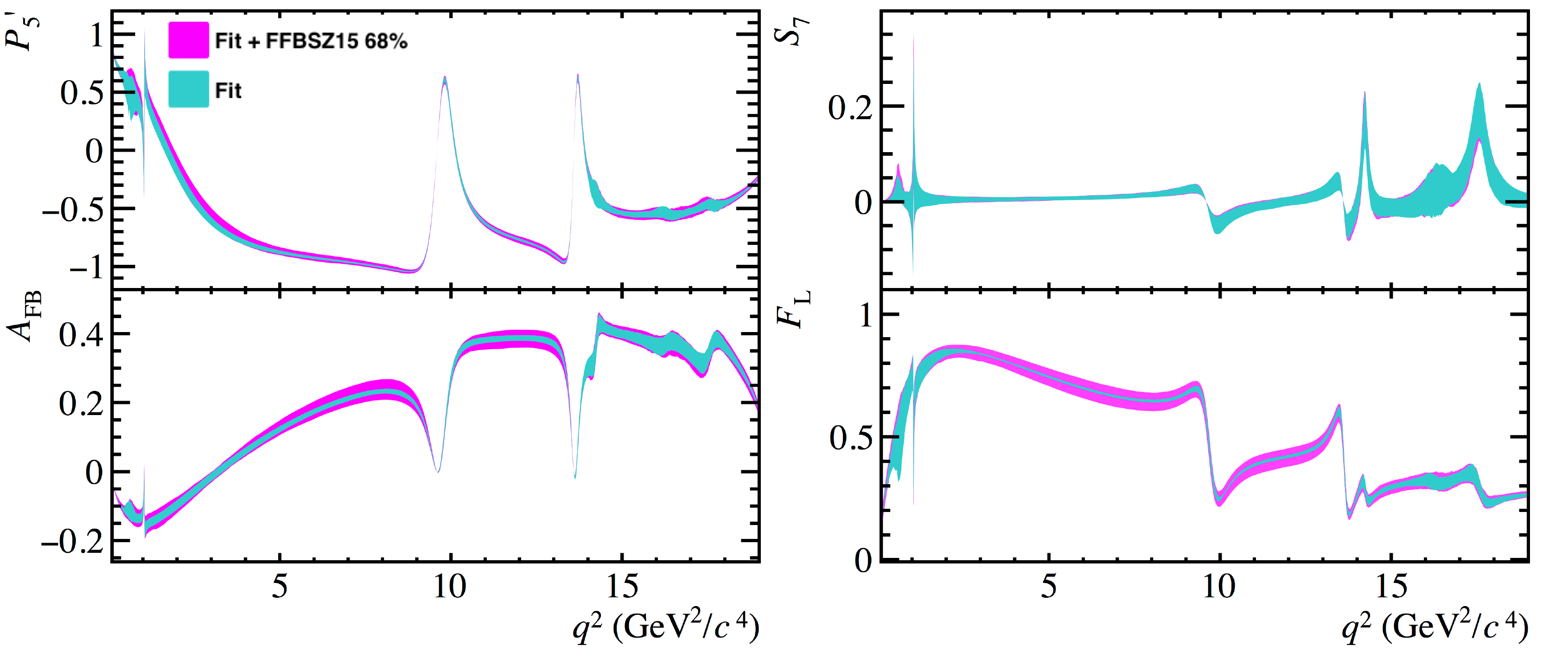}
  \caption{Predictions of the observables $P_{5}'$, $A_{\rm FB}$, $S_{7}$ and $F_{L}$ in the SM using the expected post-fit precision of the non-local parameters $\Delta C_{9\,\,\lambda}^{\rm total}$ 
    at the end of Run2 of the LHC. A sample of $\mathcal{O}(10^{6})$ simulated \BdbToKstmm decays 
    that include contributions from both short-distance and non-local components, is used 
    to determine the parameters of $\Delta C_{9\,\,\lambda}^{\rm total}$. 
    The decays are simulated in the SM, with the parameters $\theta_{j}^{0}$, $\zeta_\lambda$ 
    and $\omega_\lambda$ set to zero. The 68\% confidence intervals are 
    shown for the statistical uncertainty (cyan band) and the combination of the 
    statistical uncertainty with the $B\to K^*$ form-factor uncertainties (magenta band) 
    given in Ref.~\cite{Straub:2015ica}.
  }\label{fig:p5p_afb_post_fit}
\end{figure}

The statistical precision on the angular observables is estimated by generating values for the non-local parameters of $\Delta C_{9\,\,\lambda}^{\rm total}$, according to a multivariate Gaussian distribution centred at 
the values used to simulate the \BdbToKstmm decays, with a covariance matrix obtained from the resulting fit to the simulated data. These values are then propagated to the angular observables in order to obtain their 68\% confidence interval as a function of $q^2$. Figure~\ref{fig:p5p_afb_post_fit} shows the statistical precision to $P_{5}'$, $A_{\rm FB}$, $S_{7}$ and $F_{L}$ in the SM, where the non-local parameters are given by Table~\ref{tab:res_vals} with $\theta_{j}^{0}=0$. The equivalent plots for the remaining $\textit{CP}$-averaged observables can be found in Appendix~\ref{app:sensitivity_additional_observables}.

By the end of Run2 of the LHC, the dominant theoretical uncertainty of the angular observables in the $q^2$ region $5<q^2<14$~\gevgevcccc, will be due to the knowledge of the $B\to K^{*}$ form-factors, rather than the non-local components. Future runs of the LHC will result in an even larger number of  \BdbToKstmm decays. Therefore, it will, in a fit that combines the experimental data and the form factor uncertainties~\cite{Hurth:2017sqw}, be possible to use experimental data to further constrain Wilson Coefficients, as well as improve the precision of $B\to K^*$ form factors and non-local contributions from charm and light quark resonances.

%% file: lepton_universality.tex
\section{Hadronic resonance effects in tests of lepton universality} 
\label{sec:lnutests}

Recent tests of lepton universality in $b\to s\ell^+\ell^-$ decays
have revealed hints of non-universal new physics entering in the
dimuon Wilson Coefficient
$C_{9}^{\mu}$~\cite{Geng:2017svp,Ciuchini:2017mik,Capdevila:2017bsm,Altmannshofer:2017fio}. The
level of this potential new physics effect is compatible with the
observed anomalies in the amplitude analyses and branching fraction
measurements of $b\to s\mu^+\mu^-$ transitions. Lepton universality
tests rely on measurements such as the ratios of branching fractions
between decays with muons and electrons in the final state. The
observables $R_{K}$ and $R_{K^*}$ are defined as

\begin{equation}
  \label{eqn:rkrkst}
 \displaystyle R_{K^{(*)}}=\dfrac{\displaystyle\int_{q^{2}_{min}}^{q^{2}_{max}} \dfrac{d\Gamma(B\to
    K^{(*)}\mu^+\mu^-)}{dq^2}dq^2}{\displaystyle\int_{q^{2}_{min}}^{q^{2}_{max}} \dfrac{d\Gamma(B\to
    K^{(*)}e^+e^-)}{dq^2}dq^2}
\end{equation}

Hadronic effects in $b\to s\ell^+\ell^-$ decays are lepton
universal and observables such as $R_{K}$ and $R_{K^*}$ can be
predicted precisely in the SM, due to the cancellation of hadronic
uncertainties. Therefore, any significant deviation between
measurements and predictions of these quantities is a clear sign of
physics beyond the SM. However, in the presence of new physics effects that
enter through the Wilson Coefficient $C_{9}^{\mu}$, the cancellation
of hadronic uncertainties is no longer exact. Consequently, in order to 
determine the exact nature of any potential new physics model, an
accurate determination of the non-local contributions in
\BdbToKstmm decays is essential.

\begin{figure}[!t]
  \centering
  \includegraphics[width=0.85\textwidth]{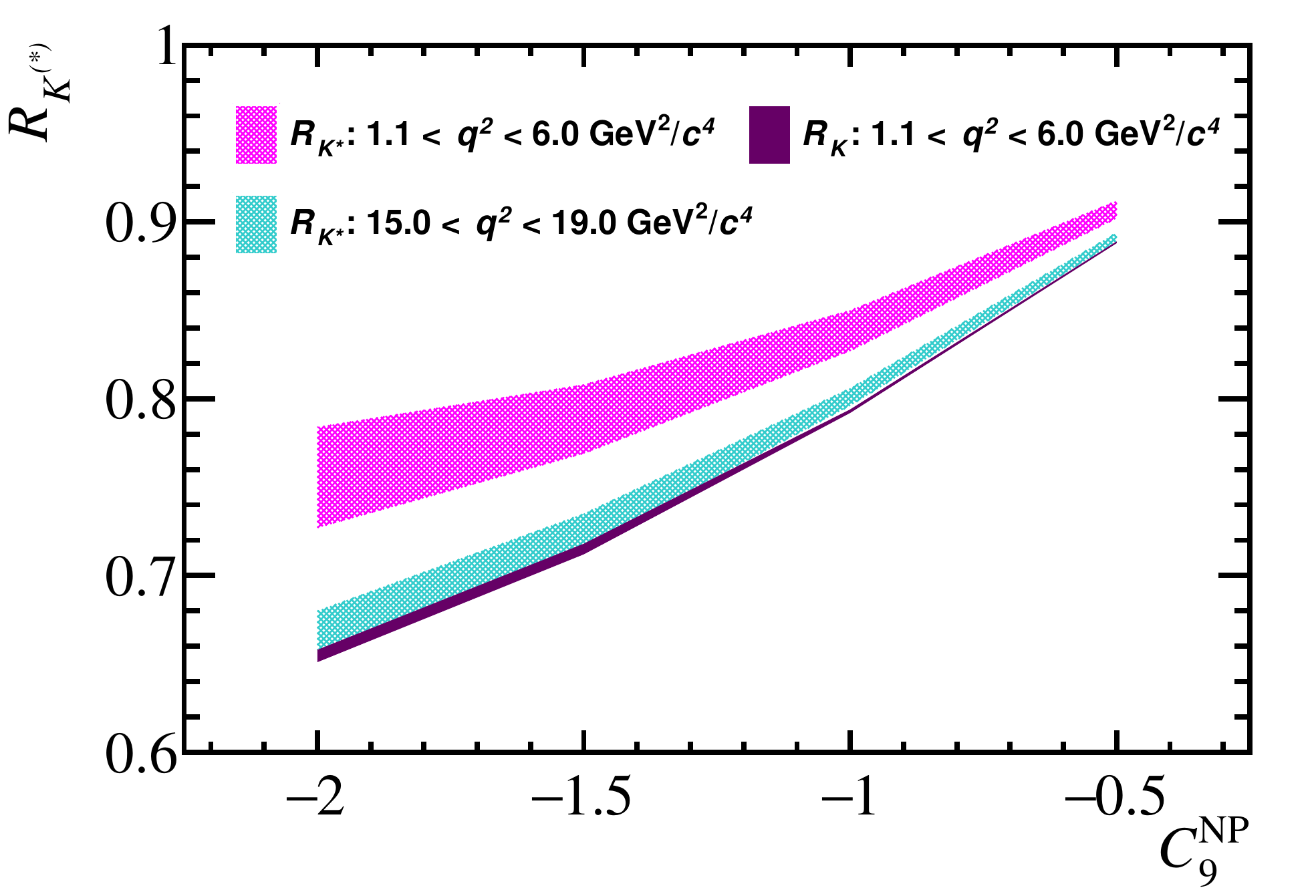}
  \caption{Predictions of $R_{K^*}$ at large recoil
    (hatched magenta) and
    low recoil (hatched cyan), and
    $R_{K}$ at large recoil (solid burgundy) for different values of
    $C_{9\,\mu}^{\rm NP}$. The $R_K$ values at low recoil are
    identical to those at large recoil and thus not shown. The interval for $R_{K^*}$
    is determined using the model described in Sec.~\ref{sec:model},
    considering the full variation of the unknown phases
    $\theta_{j}^{0}$. In contrast the 68\% confidence interval of the
    $R_{K}$ prediction is obtained using the measured non-local
    contributions in \BuToKmm decays~\cite{LHCb-PAPER-2016-045}.}
  \label{fit:rkst}
\end{figure}

The model of $\Delta C_{9}^{\rm total}$ discussed in
Sec.~\ref{sec:model} is used to provide a prediction for $R_{K^*}$
that accounts for the residual dependence on the unknown phases
$\theta_{j}^{0}$. Figure~\ref{fit:rkst} summarises this prediction in
models with values of $C_{9\,\mu}^{\rm NP}$ between -0.5 and -2.0, as
suggested by global analyses of $b\to s\mu^+\mu^-$ transitions. The
confidence interval for $R_{K^*}$ is determined by considering the
full variation of the unknown phases $\theta_{j}^{0}$. The residual
form factor uncertainty is found to be subdominant compared to the
variation of the phase. A prediction for $R_K$ is also provided,
which uses the long distance contributions measured in
Ref.~\cite{LHCb-PAPER-2016-045} with the 68\% confidence interval determined 
by treating the measured non-local parameters as
uncorrelated. It can be seen that when the experimental data is used for 
measuring the phase of the non-local contribution, the residual uncertainty becomes very small.
It is worth noting that for $C_{9\,\mu}^{\rm NP}=0$, 
there is no dependence on the unknown phase
$\theta_{j}^{0}$. Tabulated values of these predictions can be found
in Appendix~\ref{app:rkst}. In the presence of new physics entering
the Wilson coefficient $C_{9\,\mu}$, a modest variation of $R_{K^*}$
with the unknown phase $\theta_{j}^{0}$ is observed. However, this
variation is around 6 times smaller than the estimated uncertainty of $R_{K^*}$ in the presence of lepton non-universal effects suggested by Ref.~\cite{Capdevila:2017bsm}.

%% file: conclusions.tex
\section{Conclusions}
\label{sec:Conclusions}

An empirical model to describe the hadronic resonance contributions in \BdbToKstmm transitions that relies on measurements of the branching fractions and polarisation amplitudes of $\Bdb\to V\Kstarzb$ decays, is presented. For a particular choice of the relative phases between the short-distance component and the hadronic amplitudes, this model was found to be in good agreement with more formal predictions such as those of Refs.~\cite{Khodjamirian:2010vf,Bobeth:2017vxj}. The approach of this paper can naturally accommodate broad hadronic contributions from $J^{PC}=1^{--}$ states such as the $\rho^0$, the $\phi$ and charm-resonances above the open charm threshold, which can be inserted into experimental analyses of $\BdbToKstmm$ decays.

The lack of knowledge of the longitudinal phase differences between $\BdbToKstmm$ and $\Bd\to V\Kstarzb$ decays results in a larger uncertainty on the predictions of the angular observables of $\BdbToKstmm$ decays compared to current approaches. A measurement of these phases is critical as it will reduce the uncertainty in the determination of the Wilson Coefficients.

In addition, the resonant contributions to the decay provide large strong-phase differences that enhance sensitivity to CP violating effects. In this way, there is no need to rely on a time dependent analysis to a \CP eigenstate. For the method to be exploited, it is required to have a model of the strong phase differences between short- and non-local contributions to $\BdbToKstmm$ transitions as proposed here. 

In the SM, observables such as $R_{K}$ and $R_{K^*}$ are independent of hadronic uncertainties. However, in the presence of non-universal effects in \btosll transitions, these observables receive uncertainties from both the form-factor calculations and the interference between short- and non-local amplitudes. Using the models described in Ref.~\cite{LHCb-PAPER-2016-045} and in this paper, predictions for $R_{K}$ and $R_{K^*}$ are provided for various choices of the Wilson coefficient $C_{9}^{\mu}$. In order to maximise the potential of observables such as $R_{K^*}$ as a way of characterising the exact physics model behind potential lepton-universality violating effects, a measurement of the non-local contributions in \BdbToKstmm decays is crucial. The data sample that will be collected by the LHCb experiment by the end of Run2 of the LHC will allow for a simultaneous amplitude analysis of both short-distance and non-local contributions to \BdbToKstmm decays across the full $q^2$ spectrum of the decay. The model described in this paper, allows for a precise determination of both of
these components.

%% file: appendix.tex
\clearpage
{\noindent\bf\Large Appendix}

\appendix

\section{Normalisation of the {\boldmath $\eta_{j}^{\lambda}$} parameters}
\label{app:norm}
The magnitude of each resonant amplitude $\eta_{j}^{\lambda}$
appearing in Eq.~\ref{eqn:h1},~\ref{eqn:h2} and ~\ref{eqn:h3} is given by
\begin{equation}
  \label{eqn:norm}
  \begin{split}
    |\eta_{j}^{0}|^{2} &=
    \displaystyle\dfrac{f_{j}^{0}\mathcal{B}(\Bdb\to V\Kstarzb)\times\mathcal{B}(V\to\mu^+\mu^-)}
      {\displaystyle \tau_B\int
        \left|8N\frac{m_Bm_{K^*}}{\sqrt{q^2}}A_{j}^{\rm
            res}(q^2)A_{12}(q^2)\right|^{2}dq^2},\\\\
    |\eta_{j}^{\parallel}|^{2} &=   	\displaystyle\dfrac{f_{j}^{\parallel}\mathcal{B}(\Bdb\to V\Kstarzb) \times\mathcal{B}(V\to\mu^+\mu^-)}
      {\displaystyle \tau_B\int
        \left|N\sqrt{2}(m_{B}^{2}-m_{K^*}^{2})A_{j}^{\rm
            res}(q^2)\frac{A_{1}(q^2)}{m_{B}-m_{K^*}}\right|^{2}dq^2},\\\\
    |\eta_{j}^{\perp}|^{2} &= \displaystyle\dfrac{f_{j}^{\perp}\mathcal{B}(\Bdb\to V\Kstarzb) \times\mathcal{B}(V\to\mu^+\mu^-)}
      {\displaystyle \tau_B\int
        \left|N\sqrt{2\lambda}A_{j}^{\rm
            res}(q^2)\frac{V(q^2)}{m_{B}+m_{K^*}}\right|^{2}dq^2},
\end{split}
\end{equation}
where the $f_{j}^{\lambda}$ factors denote the measured polarisation fraction of
$\Bdb\to V\Kstarzb$ decays and $\tau_B$ denotes the $B$-meson lifetime.
\clearpage

\section{\textit{CP}-averaged observables}
\label{app:cp_observables}

\begin{figure}[!h]
\centering
\includegraphics[width=1\textwidth]{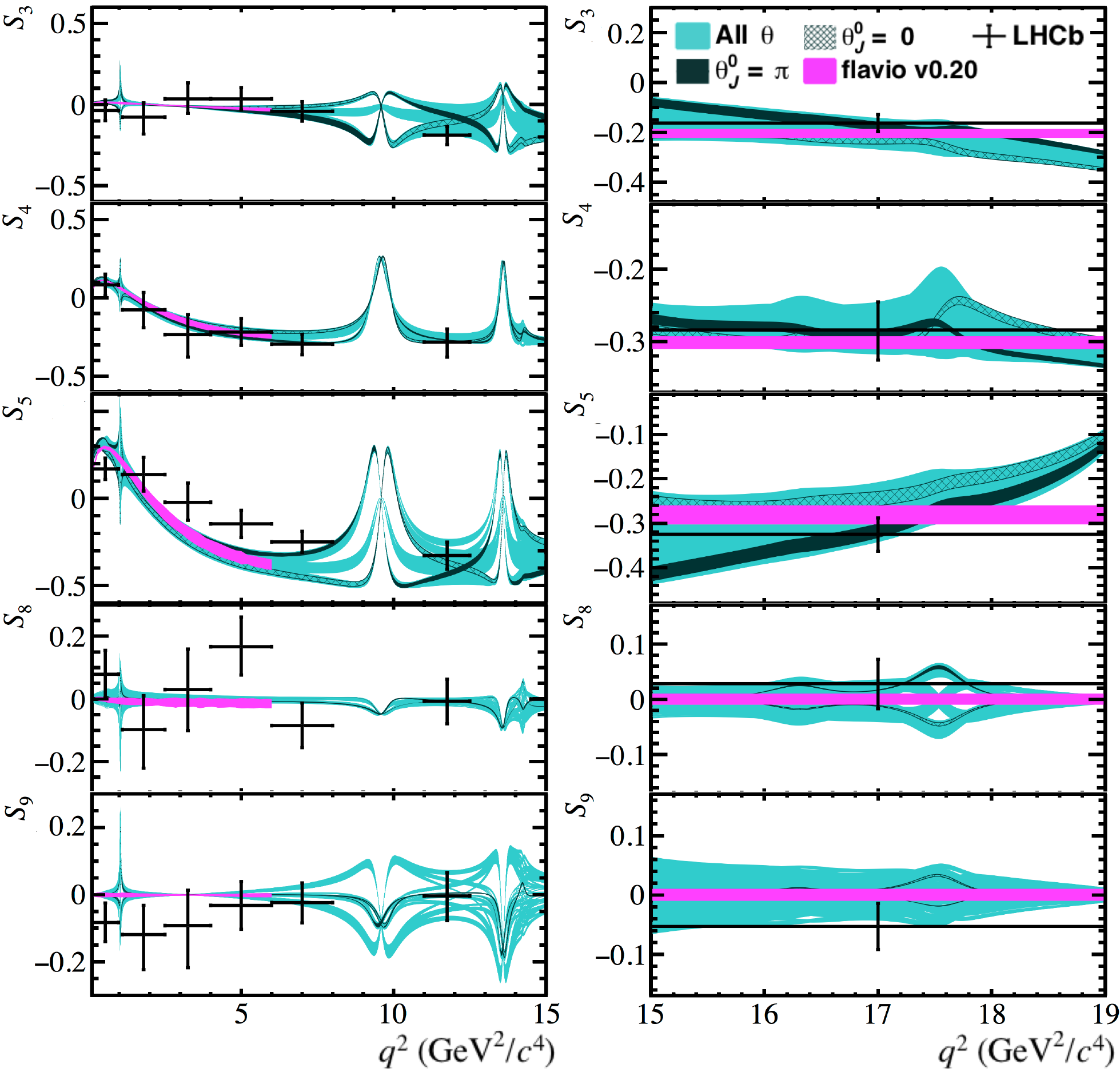}\\
\caption{Distributions of the \textit{CP}-averaged  observables in the SM as a function of $q^2$ below (left) and above (right) the open charm threshold (cyan). Specific choices are highlighted for $\theta_{j}^{0}=0$ (hatched band) and $\theta_{j}^{0}=\pi$ (dark band). The measured values of the observables from Ref.~\cite{LHCb-PAPER-2015-051} are also shown (black points). The theoretical predictions (magenta band) using \texttt{flavio}~\cite{ref:flavio} are shown for comparison.
  }
\label{fig:cp_observables}
\end{figure}

\clearpage

\clearpage

\section{Experimental sensitivity for \textit{CP}-averaged observables}
\label{app:sensitivity_additional_observables}

\begin{figure}[!h]
  \centering
  \includegraphics[width=0.45\textwidth]{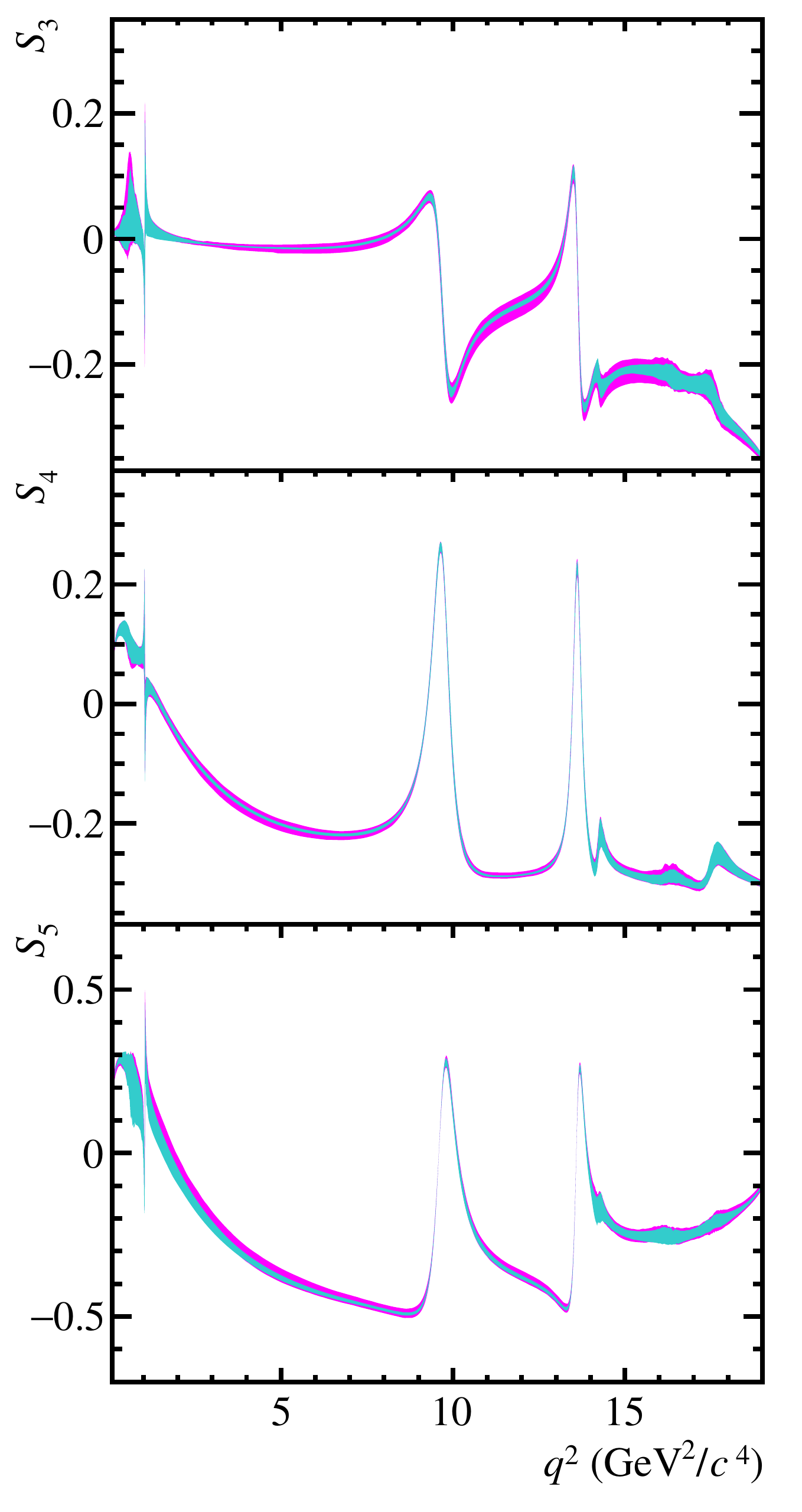}
  \includegraphics[width=0.45\textwidth]{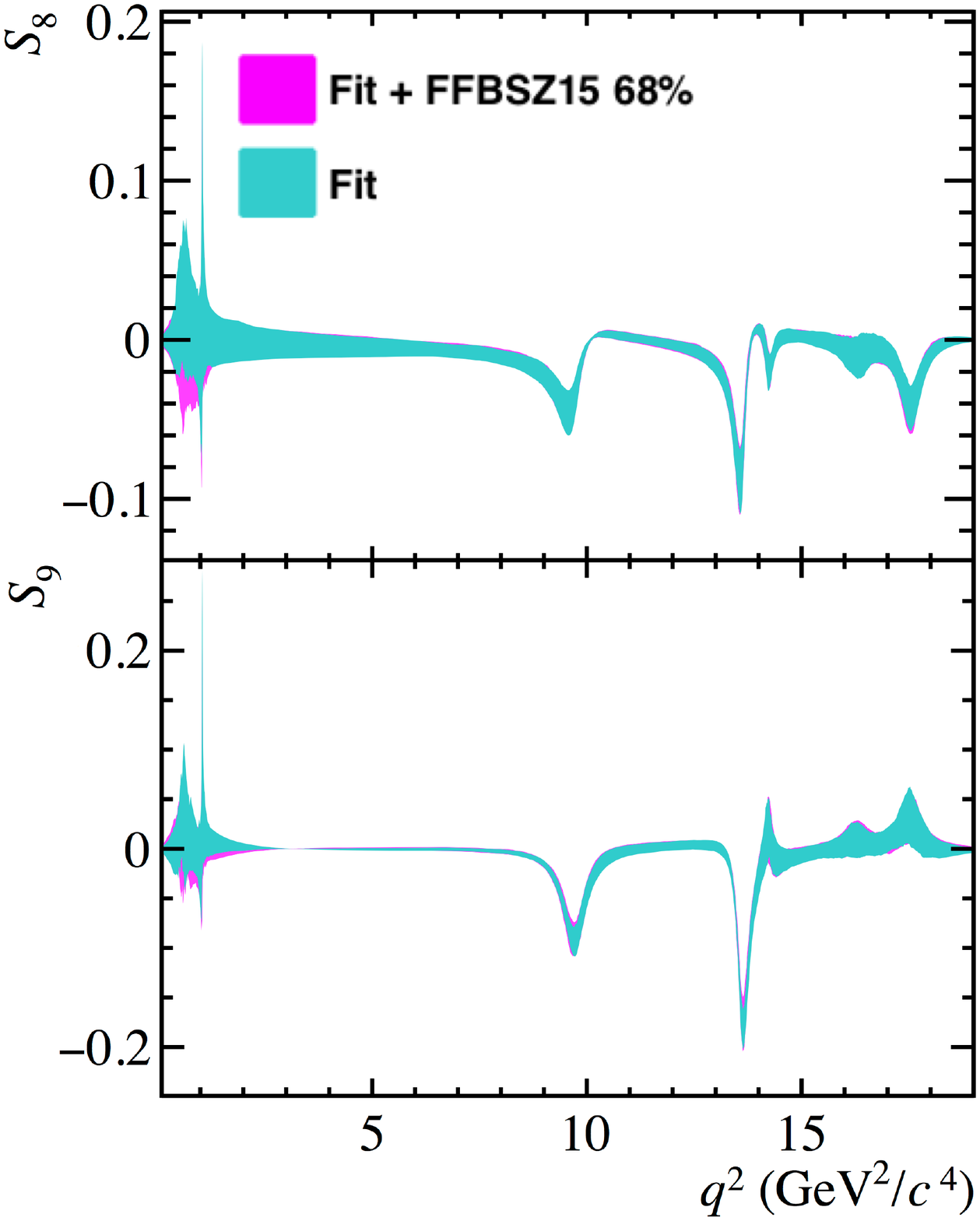}
  
  \caption{Predictions of the remaining \CP-averaged observables in the SM using the expected post-fit precision of the long-distance parameters $\Delta C_{9\,\,\lambda}^{\rm total}$ 
    at the end of Run2 of the LHC. A sample of $\mathcal{O}(10^{6})$ simulated \BdbToKstmm decays 
    that include contributions from both short- and long-distance components, is used 
    to determine the parameters of $\Delta C_{9\,\,\lambda}^{\rm total}$. 
    The decays are simulated in the SM, with the parameters $\theta_{j}^{0}$, $\zeta_\lambda$ 
    and $\omega_\lambda$ set to zero. The 68\% confidence intervals are 
    shown for the statistical uncertainty (cyan band) and the combination of the 
    statistical uncertainty with the $B\to K^*$ form-factor uncertainties (magenta band) 
    given in Ref.~\cite{Straub:2015ica}.
  }\label{fig:observables_post_fit}
\end{figure}

\section{{\boldmath $R_{K^*}$ and $R_{K}$} predictions}
\label{app:rkst}
\begin{table}[!!!h]
\centering
  \caption{Predictions of $R_{K^*}$ and $R_{K}$ at large and low
    recoil for different values of $C_{9\,\mu}^{\rm NP}$. The interval for $R_{K^*}$
    is determined using the model described in Sec.~\ref{sec:model},
    considering the full variation of the unknown phases
    $\theta_{j}^{0}$. The uncertainty due to the residual form factor
    dependence is found to be subdominant. In contrast, the 68\% confidence interval of the
    $R_{K}$ prediction is obtained using the measured long distance
    contributions in \BuToKmm decays~\cite{LHCb-PAPER-2016-045}.}
  
  \renewcommand\arraystretch{1.4} \setlength\minrowclearance{2.4pt}
\scalebox{0.85}{
\begin{tabular}{l c c c c}
  \hline
    Observable & $C_{9\,\mu}^{\rm NP}=-0.5$ & $C_{9\,\mu}^{\rm NP}=-1.0$ & $C_{9\,\mu}^{\rm NP}=-1.5$ & $C_{9\,\mu}^{\rm NP}=-2.0$ \\
    \hline
    &\multicolumn{4}{|c|}{$1.1<q^2<6.0$~\gevgevcccc} \\
    \hline
    $R_{K^*}$   & $[0.902,0.912]$ & $[0.827,0.850]$ & $[0.769,0.808]$ & $[0.727,0.784]$ \\
    $R_{K}$     & $[0.888,0.889]$        & $[0.792,0.794]$        & $[0.712,0.718]$        & $[0.651,0.658]$ \\
    \hline
    &\multicolumn{4}{|c|}{$15<q^2<19$~\gevgevcccc} \\
    \hline
    $R_{K^*}$   & $[0.889,0.894]$ & $[0.796,0.806]$ & $[0.719,0.735]$ & $[0.658,0.680]$ \\
    $R_{K}$     & $[0.888,0.889]$        & $[0.792,0.794]$        & $[0.712,0.718]$        & $[0.651,0.658]$ \\
\hline\hline
\end{tabular}
}
  \label{tab:rkst}
\end{table}